\documentclass[prb,twocolumn,superscriptaddress,showpacs]{revtex4} 

\usepackage{graphicx}
\usepackage{amsmath}
\usepackage{amssymb}
\usepackage{amsfonts}

\newcommand{\vF}{v_{\mathrm F}}
\newcommand{\bq}{{\mathbf q}}
\newcommand{\bK}{{\mathbf K}}
\newcommand{\bqD}{{\mathbf q}_{\mathrm D}}
\newcommand{\diag}{{\mathop{\rm{diag}}\nolimits\,}}
\newcommand{\bsigma}{{\boldsymbol\sigma}}

\begin{document}

\title{Resonant modes in strain-induced graphene superlattices}

\author{F. M. D. Pellegrino}
\affiliation{Dipartimento di Fisica e Astronomia, Universit\`a di Catania,
Via S. Sofia, 64, I-95123 Catania, Italy}
\affiliation{CNISM, UdR di Catania, I-95123 Catania, Italy}

\author{G. G. N. Angilella}
\affiliation{Dipartimento di Fisica e Astronomia, Universit\`a di Catania,
Via S. Sofia, 64, I-95123 Catania, Italy}
\affiliation{CNISM, UdR di Catania, I-95123 Catania, Italy}
\affiliation{Scuola Superiore di Catania, Universit\`a di Catania,
Via Valdisavoia, 9, I-95123 Catania, Italy}
\affiliation{INFN, Sezione di Catania, I-95123 Catania, Italy}

\author{R. Pucci}
\affiliation{Dipartimento di Fisica e Astronomia, Universit\`a di Catania,
Via S. Sofia, 64, I-95123 Catania, Italy}
\affiliation{CNISM, UdR di Catania, I-95123 Catania, Italy}

\begin{abstract}

We study tunneling across a strain-induced superlattice in graphene. In studying
the effect of applied strain on the low-lying Dirac-like spectrum, both a shift
of the Dirac points in reciprocal space, and a deformation of the Dirac cones is
explicitly considered. The latter corresponds to an anisotropic, possibly
non-uniform, Fermi velocity. Along with the modes with unit transmission usually
found across a single barrier, we analytically find additional resonant modes
when considering a periodic structure of several strain-induced barriers. We
also study the band-like spectrum of bound states, as a function of conserved
energy and transverse momentum. Such a strain-induced superlattice may thus
effectively work as a mode filter for transport in graphene.

\end{abstract}


\maketitle

Graphene is a single layer of carbon atoms in the $sp^2$ hybridization state,
arranged according to a honeycomb lattice \cite{Novoselov:04,Novoselov:05}.
Transport properties in graphene are largely determined by its reduced
dimensionality, which characterizes its remarkable electronic properties
\cite{CastroNeto:08,Abergel:10}. These include low-energy quasiparticles with a
Dirac-like spectrum and a linearly vanishing density of states (DOS) at the
Fermi level.  Evidence of such an unconventional behaviour is to be found in
several electronic properties, such as Klein tunneling
\cite{MiltonPereira:06,Barbier:09,Barbier:10,Barbier:10a,Peres:09}, the optical
conductivity \cite{Wang:08,Stauber:08a,Pellegrino:09b}, and the plasmon
dispersion relation \cite{Hwang:07a,Polini:09,Pellegrino:10a,Pellegrino:10c}.
These have been predicted to depend quite generally on applied strain
\cite{Pellegrino:11}, following the earlier suggestion that suitably deformed
graphene sheets could be engineered into nanodevices with the desired electron
properties \cite{Pereira:09}. For instance, it has been recently demonstrated
that the electrical properties of epitaxial graphene on SiC strongly depend on
the local strain induced in graphene by the substrate \cite{Low:12}. One thus
expects that a suitable pattern of periodically repeating stripes, with
alternating values of strain, \emph{i.e.} a strain-induced superlattice, may
produce coherent effects on single particle transport, depending on the energy
and momentum of the incident electrons. Here, we therefore study the possible
occurrence of resonant states within a strain-induced superlattice in graphene.

\begin{figure}[t]
\centering
\includegraphics[width=0.7\columnwidth]{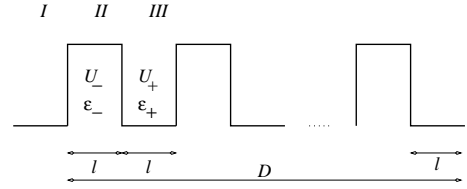}
\caption{Schematic plot of the superlattice of $N$ identical barriers, with
$\ell$ denoting both each barrier's width and the interbarrier separation, while
$D=2N\ell$. Subscript $-$ refers to the region within a barrier (labelled II),
while subscript $+$ refers to the interbarrier region (labelled I and III).}
\label{fig:scheme}
\end{figure}

We consider quasiparticle transmission across $N$ identical barriers, each of
width $\ell$, the inter-barrier separation being also $\ell$, such that $2N\ell
=D$ (Fig.~\ref{fig:scheme}). Let $x$ denote the coordinate orthogonal to the
barriers, forming an angle $\theta$ with the graphene zig-zag direction. Thus,
$\theta=0$ (\emph{resp.,} $\theta=\pi/2$) will refer to a superlattice oriented
along the zig-zag (\emph{resp.,} armchair) direction. Such a superlattice is
usually obtained via a step-wise varying gate potential $U(x)=U_\pm$, with
$U(x)= U_-$ within each barrier [$2(m-1)\ell\leq x \leq (2m-1)\ell$, $m=1,\ldots
N$], and $U(x)= U_+$ between two neighboring barriers [$(2m-1)\ell\leq x \leq
2m\ell$, $m=1,\ldots N$]. Here, we will additionally consider a nonuniform
profile of uniaxial strain $\varepsilon = \varepsilon(x)$ applied along the
$\theta$ direction, with strain modulus alternating between the values
$\varepsilon(x) = \varepsilon_\pm$ inside and outside a barrier, as above. Such
a dependence approximates a smooth periodic strain wave with period $2\ell$, as
a train of sharp steps.

Within each barrier, strain is described by the tensor
${\boldsymbol\varepsilon}=\frac{1}{2}\varepsilon[(1-\nu) + (1+\nu)A(\theta)]$,
where $\varepsilon=\varepsilon(x)$ is the strain modulus, $\nu=0.14$ is 
Poisson's ratio for graphene \cite{Farjam:09}, and
$A(\theta)=\cos(2\theta)\sigma_z + \sin(2\theta)\sigma_x$, with $\sigma_i$
($i=x,y,z$) denoting the Pauli matrices. Within a quite general tight-binding
approach \cite{CastroNeto:08}, strain then enters the electronic properties
through the dependence of the hopping parameters on the lattice vectors
\cite{Pereira:08a}. Expanding such a tight-binding Hamiltonian to linear order
in the strain modulus, one finds that the low-lying spectrum can still be
described by a Dirac-like Hamiltonian, but now (i) applied strain shifts the
location of the Dirac points in reciprocal space with respect to $\pm\bK$ at the
vertices of the first Brillouin zone, and (ii) it induces a
deformation of the Dirac cones, which can be accounted in terms of an
anisotropic Fermi velocity $\vF$. Specifically, one finds for the Hamiltonian
under applied strain
\begin{equation}
H = \hbar\vF \, {\mathcal U}^\dag (\theta) \,\tilde{\bsigma}\cdot\bq \,
{\mathcal U}(\theta) ,
\label{eq:Hamiltonian}
\end{equation}
where $\bq=(q_1,q_2)^\top$ measures the wave vector displacement from the
shifted Dirac points $\bqD a = \pm (\kappa_0
\varepsilon(1+\nu)\cos(2\theta),-\kappa_0
\varepsilon(1+\nu)\sin(2\theta))^\top$,  $\tilde{\sigma}_i = (1-\lambda_i
\varepsilon)\sigma_i$ ($i=1,2$) take into account of the strain-induced
deformation of the Fermi velocity, with $\lambda_x = 2\kappa$, $\lambda_y = -2\kappa\nu$,
${\mathcal U}(\theta) = \diag(1,e^{-i\theta})$ is the unitary matrix performing
a rotation mapping the zig~zag direction onto the the direction $x$ of applied
strain, $\kappa_0 = (a/2t)|\partial t/\partial a| \approx 1.6$ is related to the
logarithmic derivative of the nearest-neighbor hopping parameter $t$ with
respect to the lattice parameter $a$ at zero strain, and $\kappa=\kappa_0 -
\frac{1}{2}$ (cf. Ref.~\onlinecite{Pellegrino:11c}).

Since the strain superlattice is uniform along the coordinate orthogonal to the
direction of applied strain, say $y$, stationary eigenmodes will be
characterized by constant energy $E$ and transverse wave vector $k_y$. The
stationary Dirac equation associated to Eq.~(\ref{eq:Hamiltonian}), with
appropriate matching conditions for the quasiparticle spinor due to the
continuity of its associated current density at the barriers' edges, can then be
equivalently recast using the transfer matrix formalism
\cite{Titov:07,Bruus:04}. Following  Ref.~\onlinecite{Pellegrino:11}, for the
transfer matrix across the first, say, barrier in Fig.~\ref{fig:scheme}, one
finds $\mathbb{M}^{(1)} (2\ell,0) =  e^{i q_{\mathrm{D}x}^{(0)} (\varepsilon_+ +
\varepsilon_- )\ell} \tilde{\mathbb{M}}^{(1)}$, where $\bqD^{(0)} = \bqD
(\varepsilon=1)$, and $\tilde{\mathbb{M}}^{(1)}$ is a unimodular matrix, $\det
\tilde{\mathbb{M}}^{(1)} = 1$. Specifically, one obtains
\begin{subequations}
\begin{eqnarray}
\tilde{\mathbb{M}}^{(1)}_{11} &=& \lambda+i\eta ,\\
\lambda &=&
\frac{\sinh (q_- \ell)}{q_-}
\frac{\sinh (q_+ \ell)}{q_+}
(\kappa_- \kappa_+ - u_- u_+ ) \nonumber\\
\label{eq:lambda}
&&+
\cosh (q_- \ell) \cosh (q_+ \ell) ,\\
\eta &=& i \Big[
\frac{u_+ u_- -\kappa_+ \kappa_-}{ q_+ q_-} \sinh (q_- \ell ) \cosh (q_+ \ell) \nonumber\\
&&
- \sinh (q_+ \ell ) \cosh (q_- \ell) \Big],
\end{eqnarray}
\label{eq:lambdaeta_gen}
\end{subequations}
where $\lambda$ is always real, whereas $\eta$ can be real or purely imaginary,
depending $k_y$ and $E$. More compactly, one also finds
\begin{equation}
\tilde{\mathbb{M}}^{(1)}_{11} = \exp(q_+ \ell) \left[ \frac{ \kappa_+ \kappa_- - u_+ u_-}{ q_+ q_-} 
\sinh(q_- \ell )+\cosh(q_- \ell ) \right].
\label{eq:Mcompact}
\end{equation}
In Eqs.~(\ref{eq:lambdaeta_gen}) and (\ref{eq:Mcompact}), we have employed
the definitions $\kappa_\pm = (1 - \lambda_y \varepsilon_\pm) (k_y -
q_{\mathrm{D}y}^{(0)} \varepsilon_\pm )/(1 -  \lambda_x \varepsilon_\pm)$,
$u_\pm = (E - U_{\pm})/[\hbar v_F (1 - \lambda_x \varepsilon_\pm )]$, and $q_\pm
= \sqrt{\kappa_\pm^2 - u_\pm^2}$. Making use of the Chebyshev identity for the
$N$th power of a unimodular matrix \cite{Yeh:77}, for the evolution matrix
across $N$ identical barriers, one finds \cite{Pellegrino:11c}
\begin{equation}
[{\tilde{\mathbb{M}}}^{(1)}]^N_{11}
 =
\frac{\sinh (Nz)}{\sinh z} \tilde{\mathbb{M}}^{(1)}_{11}
-
\frac{\sinh((N-1)z)}{\sinh z} ,
\label{eq:cheb}
\end{equation}
where $\cosh z=\lambda$. Finally, the transmission can be related to the
evolution matrix as
\begin{equation}
T_N(E,k_y )=\left| [{\tilde{\mathbb{M}}}^{(1)}]^N_{11} \right|^{-2}.
\label{eq:transmission} 
\end{equation}

We are now in the position to discern whether an electronic mode is
characterized by an oscillating or evanescent behavior far from the barrier
superlattice. To this aim, we preliminarly observe that, depending on $E$ and
$k_y$, one has a propagating (\emph{resp.,} evanescent) wave for $q_\pm^2 <0$
(\emph{resp.,} $q_\pm^2 >0$), where the subscript $+$ refers to the region
between two consecutive barriers [$(2m-1)\ell\leq x \leq 2m\ell$, $m=1,\ldots
N$], and the subscript $-$ refers to the region within a barrier
[$2(m-1)\ell\leq x \leq (2m-1)\ell$, $m=1,\ldots N$] (Fig.~\ref{fig:scheme}). 

\begin{figure}[t]
\centering
\includegraphics[height=0.9\columnwidth,angle=-90]{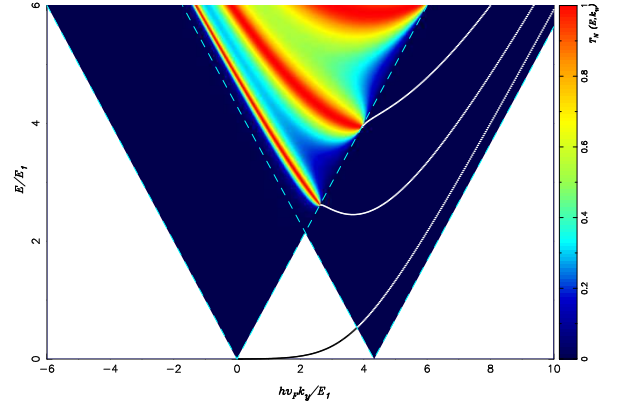}
\caption{(Color online) Single electron transmission $T_1 (E,k_y )$,
Eq.~(\ref{eq:transmission}) across a single barrier ($N=1$, $\ell=25$~nm), as a
function of scaled transverse wave vector $\hbar \vF k_y /E_1$ and scaled energy
$E/E_1$, Eq.~(\ref{eq:E1}), with $E_1 \approx 40$~meV. Here, strain is applied
along the armchair direction, $\theta=\pi/2$, and we set $\varepsilon_- = 0.02$,
$\varepsilon_+ = 0$, and $U_\pm = 0$. Cyan dashed lines delimit cones
corresponding to the (deformed) Dirac cones outside (left cone) and within
(right cone) the barrier (regions I+III and II, respectively, in
Fig.~\ref{fig:scheme}). Solid lines outside the left Dirac cone correspond to
bound modes.}
\label{fig:states1}
\end{figure}

\begin{figure}[t]
\centering
\includegraphics[height=0.9\columnwidth,angle=-90]{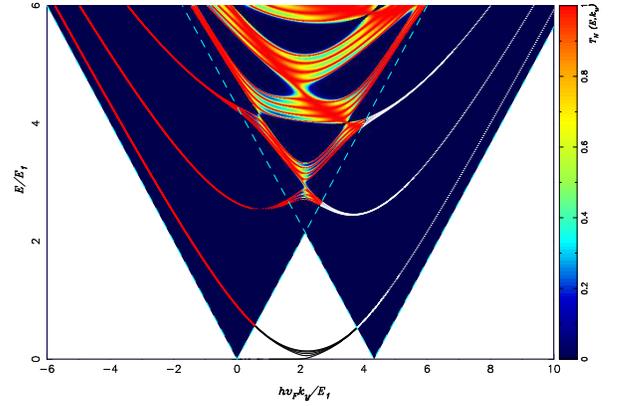}
\caption{(Color online) Single electron transmission $T_5 (E,k_y )$,
Eq.~(\ref{eq:transmission}) across a superlattice of $N=5$ identical barriers
(Fig.~\ref{fig:scheme}), as a function of scaled transverse wave vector $\hbar \vF k_y
/E_1$ and scaled energy $E/E_1$, Eq.~(\ref{eq:E1}). All other parameters are as in
Fig.~\ref{fig:states1}. Red lines outside the right cone correspond to resonant
modes.}
\label{fig:statesN}
\end{figure}

Fig.~\ref{fig:states1} shows the single electron transmission $T_N (E,k_y )$
across a single barrier, Eq.~(\ref{eq:transmission}) with $N=1$, as a function
of the transverse wave vector $\hbar\vF k_y /E_1$ and energy $E/E_1$, each
scaled by the characteristic energy
\begin{equation}
E_1 = \frac{\pi\hbar\vF}{2\ell\gamma} ,
\label{eq:E1}
\end{equation}
where $\gamma = \frac{1}{2}[ (1-\lambda_x \varepsilon_+ )^{-1} + (1-\lambda_x
\varepsilon_- )^{-1} ]$.  Here and in the following, strain is applied along the
armchair direction, $\theta=\pi/2$, and we set $\varepsilon_- = 0.02$,
$\varepsilon_+ = 0$, and $U_\pm = 0$. In Fig.~\ref{fig:states1}, cyan dashed
lines delimit the two (deformed) Dirac cones defined by $q_+^2 < 0$ (left cone)
and $q_-^2 < 0$ (right cone), corresponding to regions I+III and II in
Fig.~\ref{fig:scheme}), respectively. One finds that $T_1 (E,k_y )$ is defined
within the left cone and is
exponentially vanishing within the intersection between both cones. This
corresponds to having propagating modes in all the three regions. In this case,
resonant modes, \emph{i.e.} propagating modes with unit transmission, are
characterized by the condition for stationary waves
\begin{equation}
\tilde{q}_- \ell = m\pi,
\label{eq:resonant}
\end{equation}
where $q_- = i \tilde{q}_-$, and $m$ is an integer.

Fig.~\ref{fig:statesN} shows the single electron transmission $T_N (E,k_y )$
across a superlattice composed of five identical barriers,
Eq.~(\ref{eq:transmission}) with $N=5$. Again, nonzero values of the
transmission are to be found within the intersection of the Dirac cones
corresponding to the region inside a barrier and between two consecutive
barriers. However, at variance of the case $N=1$, because of multiple
scatterings, a nonzero transmission is also possible beyond the cone $q_-^2 <
0$. This corresponds to having evanescent modes within the barriers. Such a
phenomenon is analogous to what happens to photons propagating across a 1D
photonic crystal with alternative layers of a left-handed and a right-handed
material (1D metamaterial) \cite{Wu:03}. As for resonant modes, $T_N
(E,k_y )=1$, besides the ones given by Eq.~(\ref{eq:resonant}) regardless of
$N$, additional resonant modes are given by the condition
\begin{equation}
\lambda = \cos \left( \frac{\pi j}{N} \right) , \quad j=1,\ldots N-1,
\label{eq:N-1}
\end{equation}
where $\lambda$ is defined by Eq.~(\ref{eq:lambda}), and $|\lambda|<1$. The
latter condition implies that these resonant modes have globally propagating
behavior. Comparing Figs.~\ref{fig:states1} and \ref{fig:statesN}, one finds
that, in the domain within both Dirac cones, in addition to the resonant modes
given by Eq.~(\ref{eq:resonant}), in the case $N>1$ there exist $N-1$ new
resonant modes given by Eq.~(\ref{eq:N-1}). It should also be noted that in the
domain within the left cone but outside the second the resonant modes, which are
only given by Eq.~(\ref{eq:N-1}), are characterized by quite narrow lines in the
transmission plots.

Outside the left Dirac cone, it is still possible to find bound states
\cite{Pereira:09,RamezaniMasir:09,Barbier:10b}. Within the transfer matrix
method, these are given by the condition \cite{Bliokh:10}
$[{\tilde{\mathbb{M}}}^{(1)}]^N_{11} =0$. For $q_+^2 > 0$ one finds evanescent
modes outside the barriers, and therefore also far from the superlattice
structure. In the case $N=1$, one finds several such confined modes within the
second cone (Fig.~\ref{fig:states1}, solid lines outside the left cone), whereof
only one such mode survives in the region outside both cones. The latter is the
surface mode analyzed in Ref.~\onlinecite{Pereira:09}. In the case $N>1$
(Fig.~\ref{fig:statesN}, solid lines outside the right cone), one finds that to
each bound mode in the single barrier case there correspond exactly $N$ bound
states outside either cones. This is remindful of electron bands in solids,
where the overlap of $N$ periodically arranged atomic orbitals give rise to a
band of $N$ states.

In conclusion, we have found that a strain-induced superlattice in graphene can
accomodate additional resonant quasiparticle states, analytically characterized
by Eq.~(\ref{eq:N-1}), besides the ones usually found across a single barrier,
given by Eq.~(\ref{eq:resonant}). One finds that applied strain modifies the
kinetic part of the quasiparticle Hamiltonian, which preserves its Dirac-like
character, but around shifted and deformed Dirac cones. This can be described in
terms of a coordinate-dependent, periodic, profile of the Fermi velocity, which
produces coherent effects on the quasiparticle transmission. Specifically, we
find resonant modes with globally propagating behavior far from the
superlattice, for conserved energy and transverse momentum within the
intersection of the two deformed Dirac cones corresponding to the two
alternating strained regions. Other modes are exponentially suppressed, and we
also discuss the spectrum of bound states, which arrange themselves as `bands',
depending on the overall number of barriers making up the superlattice. We thus
surmise that a strain-induced superlattice in graphene can be used as a filter
for the resonant modes here discussed.

\begin{small} 
\bibliographystyle{apsrev}
\bibliography{a,b,c,d,e,f,g,h,i,j,k,l,m,n,o,p,q,r,s,t,u,v,w,x,y,z,zzproceedings,Angilella}
\end{small}

\end{document}